# Power and Performance Efficient SDN-Enabled Fog Architecture

Adnan Akhunzada, Senior Member, IEEE; Sherali Zeadally, Senior Member, IEEE; Saif ul Islam

**Abstract**—Software Defined Networks (SDNs) have dramatically simplified network management. However, enabling pure SDNs to respond in real-time while handling massive amounts of data still remains a challenging task. In contrast, fog computing has strong potential to serve large surges of data in real-time. SDN control plane enables innovation, and greatly simplifies network operations and management thereby providing a promising solution to implement energy and performance aware SDN-enabled fog computing. Besides, power efficiency and performance evaluation in SDN-enabled fog computing is an area that has not yet been fully explored by the research community. We present a novel SDN-enabled fog architecture to improve power efficacy and performance by leveraging cooperative and non-cooperative policy-based computing. Preliminary results from extensive simulation demonstrate an improvement in the power utilization as well as the overall performance (i.e., processing time, response time). Finally, we discuss several open research issues that need further investigation in the future.

## 1. INTRODUCTION

Software Defined Networks (SDNs) have strong potential to meet the imminent demands and requirements of next-generation networks. SDNs have revolutionized the networking industry with its promising architecture that supports network programmability. The constantly evolving digital landscape and users' traffic requirements for smart connected communities require scalable, reliable, secure, and robust infrastructures. Despite the recent advances in Information and Communication Technologies (ICTs) along with various underlying architectures, there is still a significant need for more computational capabilities to meet future challenges [1]. Besides, the pervasive nature of IoT devices has not only imposed a heavy burden on communication bandwidth but has also resulted in high transmission latency and degraded both the Quality-of-Service (QoS) and Quality-of-Experience (QoE) of end users [2]. Besides, interactive communications, increased delays with moving users/objects, large data flows, support for mobility and geo-distribution have opened up new challenges for the next-generation networks.

Moreover, the capability to respond in real-time while handling large amounts of data is a challenge in pure SDNs. To handle a lot of data, over the years, diverse computing paradigms have been used including supercomputing, accelerator-based computing, grid computing, cluster computing, and cloud computing. To address the real-time requirements of emerging applications, recently fog computing has emerged, and this new paradigm has been receiving considerable attention from both industry and the academic research community [2]. Furthermore, researchers have addressed the significance of adding a fog layer in the context of various architectures such as the Internet of Things (IoT) to potentially evaluate performance and energy trade-offs and mobility support for fog computing in SDNs. Consequently, there is a need to analyze the performance and power consumption in SDN-enabled fog computing.

To the best of our knowledge, this is the first effort to analyze power efficiency and performance in SDN-enabled fog architecture using cooperative and non-cooperative fog computing policies that is detailed in Section 2. However, our previously published work [12, 13] comprehensively elaborates QoS service provisioning in fog computing.

The main contributions of this work are as follows:

- We propose a novel SDN-enabled fog computing architecture that leverages cooperative and non-cooperative policy-based computing to improve power efficiency and QoS performance.
- Our preliminary results demonstrate the efficacy of the proposed power and performance efficient SDN-enabled fog architecture.
- Finally, we identify some future research challenges in integrated SDN and fog environments that must be addressed in the future.

The remainder of this paper is organized as follows. Section II presents an overview of our proposed SDN-enabled fog architecture. Section III presents performance evaluation and results and discussions. Section IV discusses open research issues and challenges. Section V concludes the paper.

## 2. SDN-enabled FOG ARCHITECURES

The section elaborates recently proposed SDN-enabled fog architectures together with our proposed architectures.

## 2.1 Overview of existing SDN-enabled fog computing architectures

The authors of [3] proposed a novel SDN-enabled mobile fog computing architecture that decouples the data plane into mobility control and data forwarding. The architecture contributes to route optimization to efficiently address performance gain in data communication and system overhead. Additionally, they also developed efficient signaling operations to provide seamless and transparent mobility support to mobile users. In [4], the authors proposed an SDN-enabled Fog Computing IoT architecture to support applications with mobility and delay-tolerance requirements. The authors of [5] proposed an SDN-enabled distributed fog computing architecture for blockchains where distributed fog computing entities allow the deployment of fog services at the edge of the IoT network [5]. In [6], the authors proposed TelcoFog, an SDN-enabled fog and cloud computing architecture for 5G networks that operate in hierarchical (i.e., parent/child or in a peer architecture for the optimal allocation of resources across distributed TelcoFog nodes).

In [7], the authors proposed a novel SDN-enabled fog computing based vehicular network (SDFC-VeNET). The architecture has three layers to provide high flexibility and scalability to the underlying vehicular networks. Further, in [8], the authors proposed a Vehicular Software-Defined Networking (i.e., VSDN) architecture where the cloud orchestrates the fog nodes in a centralized fashion. VSDN enables content dissemination to efficiently accommodate large number of vehicular users with various kinds of communication technologies and devices. In [9], the authors proposed Fog-enabled Software-Defined Networking Vehicular Ad-hoc Network (FSDN-VANET) architecture to enable mobile operators to offer their clients valuable transportation services.

Our proposed architecture differs from the previously proposed SDN-enabled fog architectures in the following ways: a) The extant architectures are either application or architecture specific. Conversely, our proposed centralized architecture is more generic, and it can be applied to various IoT/ IoE environments, b) Fog computing mainly uses cooperative or non-cooperative mechanisms to handle IoT requests [12]. Hence, we selected cooperative and non-cooperative policies for the evaluation of our proposed architecture, c) Our policy-based SDN-enabled fog computing architecture provides centralized control management facility that makes it a promising architecture for emerging large scale distributed networks, and d) Our devised centralized policy based computing architecture opens up new opportunities to implement various centralized policies such as pricing, ownership, conflict of interest, and so on for distributed cloud/fog/edge computing.

## 2.2 Proposed SDN-enabled fog architecture

Figure 1 depicts our proposed SDN-enabled fog computing architecture. Besides, a detailed architectural overview can be seen in our previous published works [9, 14]. The programmability aspects of the control plane allow users to implement, customize, and program network functionalities such as topological management, network orchestration and abstraction, various types of service mapping, intrusion detection and prevention systems. Thus, cooperative and non-cooperative computing policies can be customized and implemented as part of an extended module of any commercial SDN controller such as Floodlight, POX, and OpenDayLight [10] to better manage the underlying network and various network functionalities at the data plane [9, 14]. The underlying SDN agents at the data plane comprises fog nodes together with other smart devices such as intelligent vehicles. Fog nodes equipped with various smart devices at the data plane are either fixed or mobile and they can upload collected data to fog servers for subsequent processing. For instance, various fog nodes can execute upload/download location-aware services. These fog devices are connected to SDN switches through the SDN-enabled wireless access infrastructure as shown in Figure 1, wherein the forwarding behavior and the control logic are controlled by the SDN based fog controller which also leverages cooperative and non-cooperative computing policies. This enables our proposed architecture to provide a centralized control management facility at the control plane and a promising solution to implement power and performance efficient SDN-enabled fog architecture.

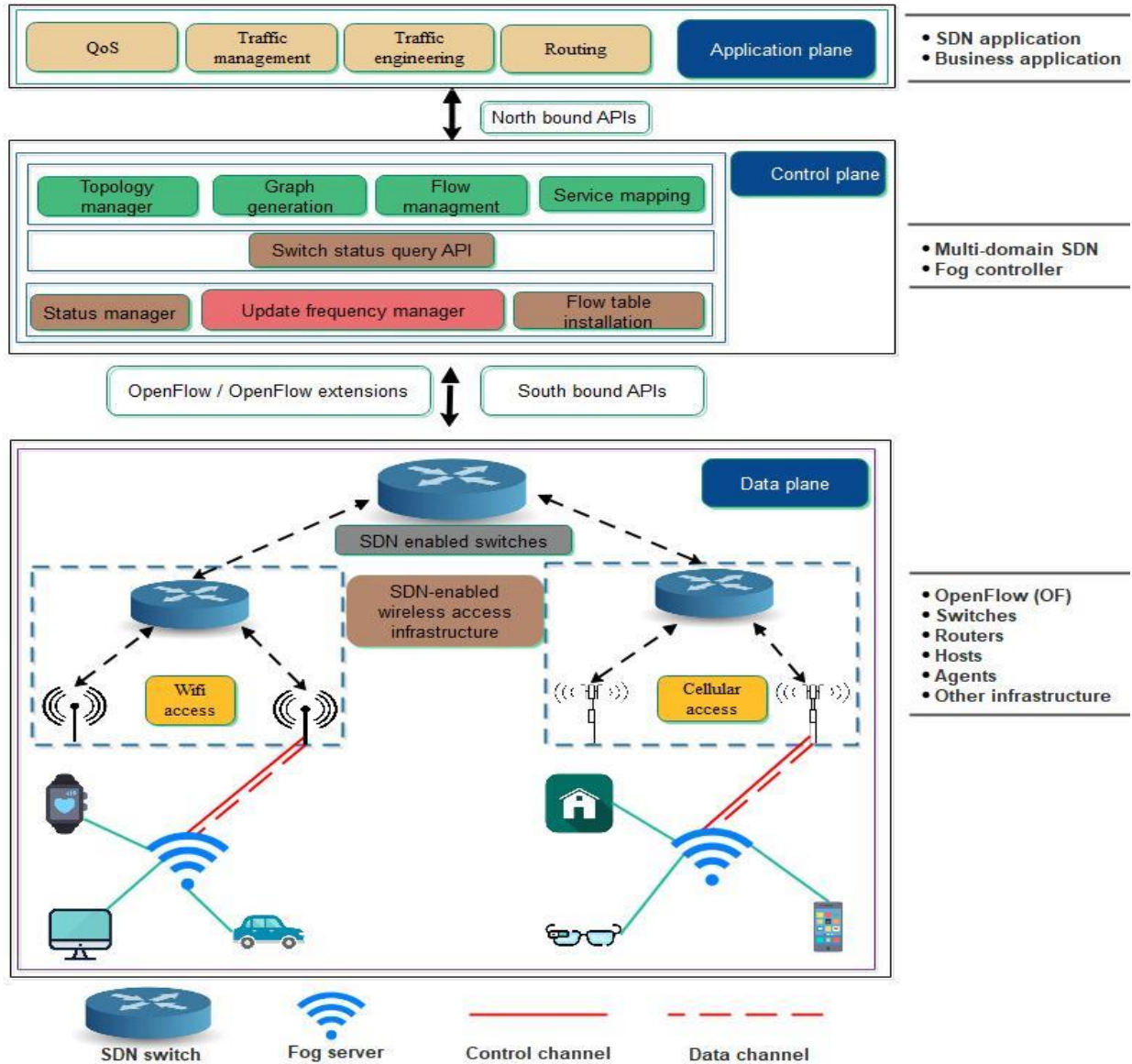

Fig. 1: A simplified view of the SDN-enabled fog computing architecture

### 2.2.1 Cooperative fog computing
In a fog computing environment, sometimes a fog server is used to its full capacity. In this scenario, there can be two options: let the incoming job wait until the occupied fog resource is released or the job should be redirected to another available fog server. In cooperative fog computing, such jobs are forwarded from one fog node to another fog node whenever and wherever needed.

### 2.2.2 Non-Cooperative fog computing
In non-cooperative fog computing, the users' jobs are directed to the nearest fog server. If the fog server has the available resources, then the request is served accordingly. In this case, if the nearest fog server is being used to its full capacity, then the request has to wait in the queue until that particular occupied resource is released. In non-cooperative fog computing, if a fog server has a large number of IoT nodes, the queuing delay of an IoT request to be served at a fog server can end up placing many tasks in the waiting queue of the fog server thereby increasing the overall delay.

### 2.2.3 Fog server selection
Selection of a fog server is crucial to assure the QoS required. In the proposed architecture, the selection of a fog server is based on the geographical proximity of an IoT device and the corresponding fog server. A set of IoT devices are associated with a nearby fog server; these devices are responsible for processing their jobs. The fog nodes and IoT devices are randomly deployed. Hence, the number of IoT devices associated with fog nodes may vary. When a request arrives at a fog server, the required resources are identified. If the fog server has no capacity to serve the newly arrived job requests, then these incoming jobs are placed in its queue. When a resource is released, the jobs in the queue are assigned to the resources accordingly. This job assignment is usually performed based on some implemented queuing policy.

## 3. EXPERIMENTAL SETUP, MEASUREMENT PROCEDURE, AND PERFORMANCE RESULTS

The section details experimental setup, evaluation metrics and performance results obtained.

### 3.1 Experimental setup

We performed the simulation tests on iFogSim [10], Amazon S3 service and Azure Cloud Service. However, the dataset [11] includes various types of synthetic data, i.e., the IoT job could contain multimedia data (audio and video), small textual (word, excel, PowerPoint etc.), and data related to the medical field. In particular, for this simulation, we have divided our data set into two main categories: a) Thick data type (large and bulky data, requiring a lot of resources in order to be processed) and b) Thin data type (not bulky, small textual data which requires minimum resources to be processed). The network topology of the targeted test case involves a cloud data center at the highest layer followed by the Proxy Server in the second layer. The fog devices are introduced at the third layer comprising fog nodes such as servers and are added depending on the demand of IoT applications. IoT devices such as cameras, smartphones, and biosensors are attached to the fog nodes and they generate various types of data aforementioned. The IoT devices and corresponding fog servers are distributed randomly at different geographical locations, whereas; the cloud services are located in Singapore and the United States and they are mainly equipped with various specifications of HP Proliant G4.

### 3.2 Evaluation metrics and measurement procedure

We conducted our performance evaluation tests using three performance metrics namely, processing time, response time, and power consumption. Processing time is the time taken (milliseconds) for a request to be processed at a fog server. Response time is the time (milliseconds) between sending a request and receiving a response from the fog server that includes various types of delays, such as propagation, queuing, and processing delays (in milliseconds).

Power consumption (Watts) is the amount of power consumed by a fog server while serving the IoT jobs. The power value is calculated based on the Millions of Instructions per Second (MIPS) required by a job and the available MIPS on a fog server. We considered these performance metrics under predefined cooperative and non-cooperative fog computing policies. We have run a simulation test 50 times for each experimentation for the performance metrics mentioned above. We collected the data for each test, and we calculated the aggregated average power consumption to explicitly show results Figure 3 shows.

### 3.3 Performance results and discussion

In this section we present the performance results obtained and discuss their significance.

#### *3.3.1 Processing time*

Figure 2 shows the processing time of IoT jobs on the fog servers. In the case of non-cooperative fog, we obtain a higher processing time compared to cooperative policy-based computing. For the non-cooperative scenario, if a resource is already reserved, then the request has to wait until the resource is released. Consequently, the request is queued at a fog server which increases the processing delay. Besides, the processing time decreases when there is cooperation among fog resources where servers cooperate with each other to satisfy the user's request. It improves the system performance in the fog computing environment thereby resulting in reduced processing overhead.

#### *3.3.2 Response Time*

Response time is an important metric to evaluate the quality of services. Figure 2 shows the mean response time of requests in a fog computing environment. Cooperative fog yields better results as compared to non-cooperative fog. When there is cooperation among fog resources, we have reduced queuing delays which minimize the response times. However, cooperation also causes an increase in the propagation delay along with a higher redirection overhead. In a non-cooperative environment, the probability of direct resource allocation is lower particularly during high- traffic load which causes an increase in queuing delays while waiting for fog resources. Consequently, the job must wait to get resources at fog servers resulting in higher queuing delays and response times.

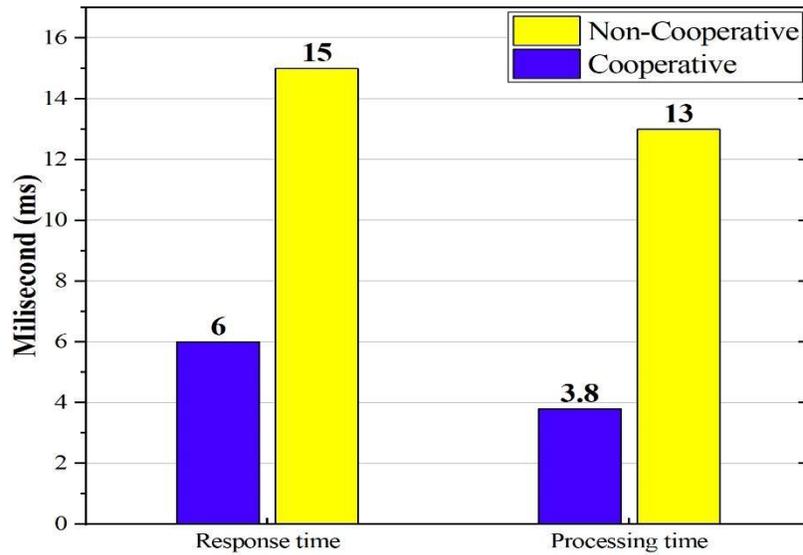
Fig. 2: Processing and response time of cooperative versus non-cooperative fog.

### *3.3.3 Average power consumption*
Figure 3 presents the average power consumption of fog servers for cooperative and non-cooperative strategies. It shows that cooperative fog performs better in terms of power consumption. The non-cooperative strategy increases the processing overhead on fog servers which also increases the server utilization. The utilization of servers is directly proportional to its dynamic power consumption as Figure 3 shows. Hence, a non-cooperative policy increases the power consumption in a fog system. In contrast, cooperative fog balances the load among fog servers and optimizes the power consumption.

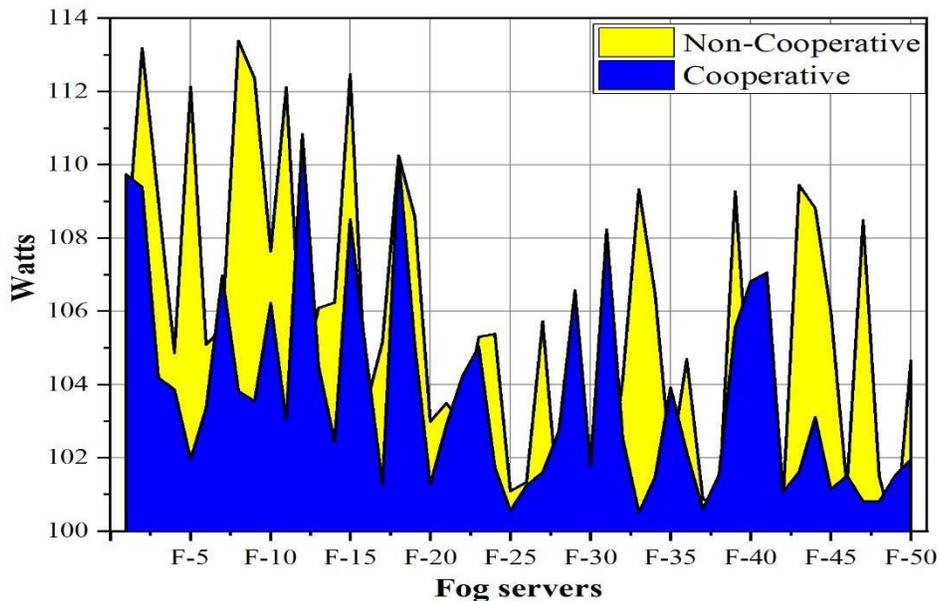
Fig. 3: Average power consumption of cooperative versus non-cooperative fog.

## 4. OPEN ISSUES AND CHALLENGES IN FOG BASED SDN COMPUTING

This section briefly explores some of the critical challenging issues. Resource management is considered as one of the most challenging tasks in distributed systems such as fog computing. However, the integration of SDN based fog computing needs serious research efforts to explore resource management issues comprehensively. On the contrary, in a fog based SDN computing environment, ensuring security and privacy becomes even more challenging because we need to consider threats at the fog infrastructure as well as threats at the network level simultaneously. Moreover, prioritizing the users' tasks, based on their requirements, for processing on particular fog nodes is an important issue that may create conflict of interest among various users. Similarly, the emergence of the novel computing paradigms such as continuum computing [15], Multi-access Edge Computing (MEC) [16], and Device-to-Device (D2D) communications and computing opens up new challenges to provide seamless communication and QoS-aware computing services where mobility of nodes is also a major concern [17, 18]. Conflict of interest issues need to be addressed to achieve smooth and efficient provision of various services. Furthermore, unlike cloud, fog based SDN computing devices may have multiple owners and the involvement of several entities and their corresponding level of communication require the development of a systematic and transparent pricing models acceptable to all stakeholders.

## 5. CONCLUSION

Recent advances in user traffic requirements and the ability to provide real-time responses while handling large amounts of data remain a challenge in pure SDNs. In contrast, fog computing has emerged to provide timely real-time responses. Keeping in view the programmability aspects of the SDN control plane, we have proposed a power and performance efficient SDN-enabled fog computing architecture. This architecture comprises cooperative and non-cooperative based policy computing to support a power and performance aware centralized control management facility for decentralized fog computing. The preliminary results we have obtained so far appear promising and could be used as a foundation to further explore performance and power issues in the new realm of SDN-enabled fog computing. Finally, we have also identified and discussed several open issues and future research directions that require further investigations.

# Biographies:

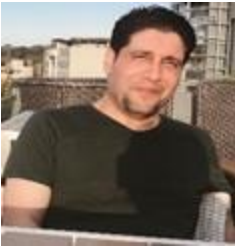

Adnan Akhunzada is a **Senior Member, IEEE** and a **Professional member, ACM** with extensive 13 years of Research and Development (R&D) experience both in ICT industry and academia. He is a cybersecurity research specialist with a proven track record of high impact published research, innovations (i.e., US Patents), EU funding's and commercial ICT products. He is currently working with Technical University of Denmark.

Sherali Zeadally earned his bachelor's degree in computer science from the University of Cambridge, England. He also received a doctoral degree in computer science from the University of Buckingham, England, followed by postdoctoral research at the University of Southern California, Los Angeles, CA. He is currently an Associate Professor at the University of Kentucky. He is a Fellow of the British Computer Society and the Institution of Engineering Technology, England.

Saif ul Islam has earned his PhD degree in computer science from Université Toulouse III Paul Sabatier, Toulouse, France, in 2015. He is currently an Assistant Professor with the Department of Computer Science, Dr. A. Q. Khan Institute of Computer Science and Information technology, Rawalpindi, Pakistan. He is a well-known researcher in the field of large-scale distributed systems.